\documentclass[a4paper,twoside,12pt]{article}

\usepackage{psfig}
\usepackage{amsfonts}
\usepackage{amssymb}

\newcommand{\ZZ}{\mathbb{Z}}

\title{More on core instabilities of magnetic monopoles}
\author{J.Striet$^1$ and F.A.Bais$^2$\\[2mm]
Institute for Theoretical Physics\\University of Amsterdam \\
Valckenierstraat 65\\1018XE Amsterdam\\The Netherlands\date{April,
2003}}

\begin{document}
\maketitle
\footnotetext[1]{jelpers@science.uva.nl}
\footnotetext[2]{bais@science.uva.nl}


\begin{abstract}
\noindent In this paper we present new results on the core
instability of the 't~Hooft Polyakov monopoles we reported on
before. This instability, where the spherical core decays in a
toroidal one, typically occurs in models in which charge conjugation
is gauged. In this paper we also discuss a third conceivable
configuration denoted as ``split core'', which brings us to some
details of the numerical methods we employed. We argue that a core
instability of 't~Hooft Polyakov type monopoles is quite a generic
feature of models with charged Higgs particles.
\end{abstract}

\section{Introduction}
Since the pioneering work of 't~Hooft and Polyakov
\cite{'thooft,polyakov} magnetic monopoles have been studied in detail
in many different models. In this paper we report on the results of
investigations following our earlier work \cite{jelper2} on the core
instability of the fundamental spherically symmetric magnetic monopole
solution. We determine the regions in parameter space where this
instability occurs and present some details of the numerical
simulations we performed. In \cite{jelper2} it was shown that in a
rather simple model the spherically magnetic monopole solution is not
the global minimal energy solution for all values of the parameters of
the model. The fact that the core topology is not fixed by the
boundary conditions at infinity and different core topologies can be
deformed into each other was already established earlier
\cite{bais2}. As we will indicate, alice theories have a special
topological feature which makes it plausible that such a core
deformation really can be favored energetically. We will also argue
that a core deformation is typically energetically favored in models
where the charged Higgs particles are light compared to the neutral
Higgs.

This paper is organized as follows. We start with a brief recap of
alice electrodynamics (AED) and its features which lead us to expect
the core meta-/instability of the monopole. Next we introduce the
specific model and discuss in some detail the numerical simulations we
performed and present our the main results. We end the paper with
conclusions and an outlook.

\section{Alice electrodynamics}
Alice electrodynamics (AED) is a gauge theory with gauge group
$H=U(1)\ltimes\ZZ_2\sim O(2)$, in a certain sense the minimally
non-abelian extension of ordinary electrodynamics. The nontrivial
$\ZZ_2$ transformation reverses the direction of the electric and
magnetic fields and the sign of the charges. In other words, alice
electrodynamics is a theory in which charge conjugation symmetry is
gauged. However, as this non-abelian extension is discrete, it only
affects electrodynamics through certain global (topological) features,
such as the appearance of alice fluxes and cheshire charges
\cite{schwarz,alford}. In this paper the cheshire phenomenon is of
great importance to us as it supports the possible core
meta-/instability of the spherically symmetric magnetic monopole
solution. The topological structure of $U(1)\ltimes\ZZ_2$ differs from
that of $U(1)$ in a few subtle points. AED allows topologically stable
localized fluxes since $\Pi_0(U(1)\ltimes\ZZ_2) = \ZZ_2$, the so
called alice fluxes. Note that in this theory this flux is
co\"existing with the unbroken $U(1)$ of electromagnetism and is
therefore not an ordinary ``magnetic'' flux. Just as a $U(1)$ gauge
theory, AED may contain magnetic monopoles, which follows from the
fact that $\Pi_1(U(1)\ltimes\ZZ_2) = \ZZ$. We note however, that due
to the fact that the $\ZZ_2$ and the $U(1)$ part of the gauge group do
not commute, magnetic charges of opposite sign belong to the same
topological sector.

It was pointed out long ago that there are interesting issues
concerning the core stability of magnetic monopoles. Fixing the
asymptotics of the Higgs field, the core (i.e., the zeros of the Higgs
field) may have different topologies, notably that of a ``ring''
rather than the conventional ``point''\footnote{In fact in AED the
Higgs field can typically only be represented by a director field if
it is in the vacuum manifold. Thus the Higgs field need not go to zero
inside the core of a defect. However for the spherically symmetric
magnetic monopole solution the Higgs field does go to zero, but for
the alice loop solution it does not.}. These core topologies are {\it
cobordant}, i.e., they can be smoothly deformed into each other and it
is a question of energetics what will be the lowest energy monopole
state \cite{bais2}. In AED such a core deformation would be
accompanied by the rather unusual delocalized version of (magnetic)
charge, the so called {\it magnetic cheshire charge}. Cheshire charge
is a key feature of AED and is a general phenomena in field theories
with (topologically) stable fluxes which are not elements of the
center of the unbroken gauge group.

\begin{figure}[!htb]
\begin{center}
\makebox[7.9cm]{\psfig{figure=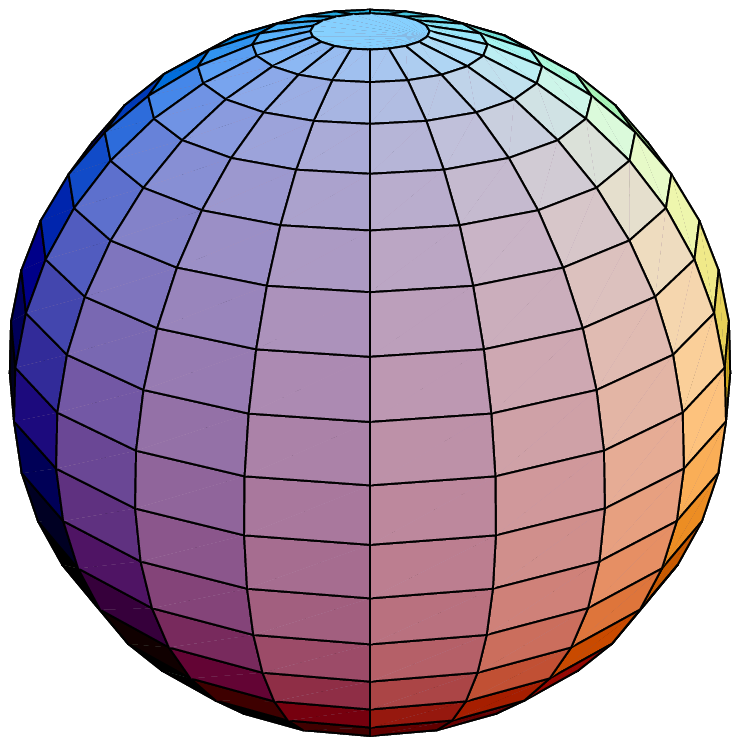,width=5cm,angle=0}}
\makebox[7.9cm]{\psfig{figure=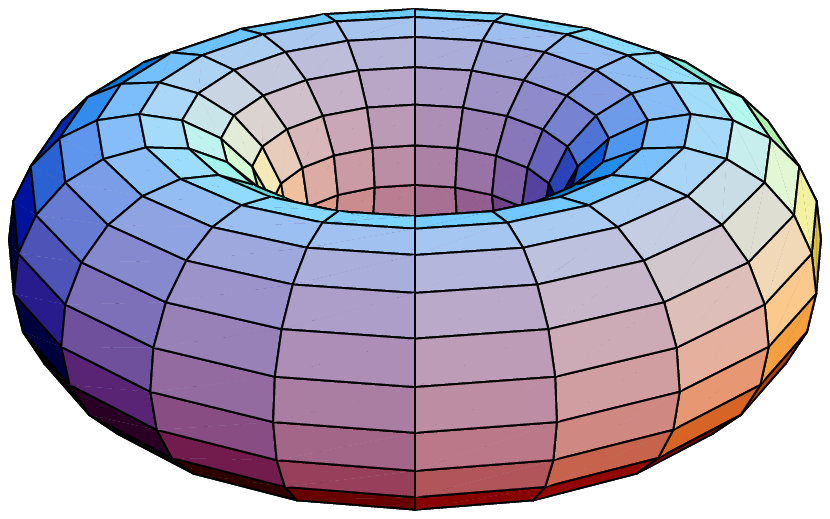,width=6.5cm,angle=0}}
\makebox[7.9cm][c]{\footnotesize{(a)}}
\makebox[7.9cm][c]{\footnotesize{(b)}}
\caption[somethingelse]{\footnotesize These figures show the two
different core structures of a magnetic defect which are naturally
present in AED. Figure (a) represents the spherically symmetric
magnetic monopole, the 't~Hooft Polyakov type monopole, and figure (b)
represents the magnetically, cheshire charged alice ring.}
\label{corestructure}
\end{center}
\end{figure}

In the specific AED model we consider the Higgs field is a symmetric
tensor, whose vacuum expectation value may be depicted as a
bidirectional arrow. The head-tail symmetry of the order parameter
reflects the charge conjugation symmetry of the theory. In AED we can
``punch a hole'' in the spherically symmetric monopole and deform it
into an alice ring, this configuration is consistent with the
continuity requirement on the order parameter because of its head-tail
symmetry. In figure \ref{corestructure} we plotted the two different
core regions one expects to find for a magnetic monopole in AED. In
this paper we determine in what part of the parameter space of the
model the monopole is meta-/unstable and where we expect the
spherically symmetric monopole to compete with a magnetically,
cheshire charged alice ring. Figure \ref{corestructure}(a) represents
the spherically symmetric magnetic monopole and figure
\ref{corestructure}(b) represents the magnetically cheshire charge
alice ring. The fact that the core of the defect can really deform
into a torus is due to the head-tail symmetry of the Higgs field in
the broken phase. We note that Higgs field only rotates over an angle
$\pi$ when going around a single flux. This is the hallmark for an
alice flux, so the core deformed spherical monopole is in fact an
alice ring carrying a magnetic cheshire charge, see also
\cite{jelper2}.

We just showed that the punctured monopole is a magnetically cheshire
charged alice loop. To understand that such a configuration can indeed
be stable we make an estimate for the energy of such a charged alice
loop. We approximate the energy of the circular alice loop of radius
$R$, by $E_{loop} = 2\pi R {\cal E}_{flux}$, with ${\cal E}_{flux}$
the energy per unit length of the alice flux, and approximate the
energy of the cheshire charge, $E_{ches}$, by a uniformly charged disc
with radius $R$. The magnetic field of the latter is given by:
\begin{eqnarray}
\vec{B}&=& \frac{-Q}{\pi R^2} \vec{\nabla} \int_0^1 \int_0^{2 \pi}
\frac{r'~ dr' d\theta'}{\sqrt{r^2+z^2 +r'^2-2rr'\cos{\theta'}}}\\ &
\equiv & \frac{Q}{R^2} ~ \vec{a}(r,z)
\end{eqnarray}
Where we have rescaled the coordinates by a factor of $R$ and $Q$ is
the total charge. The field energy is then given by:
\begin{eqnarray}
E_{ches} &=& \int \vec{B} \cdot \vec{B}~ R^3 d^3x\\ &=& \frac{Q^2}{R}
\int \vec{a}(r,z) \cdot \vec{a}(r,z)~ d^3x\\ &\equiv& \frac{Q^2}{R}~ A
\end{eqnarray}
Where $A$ is a dimensionless constant determined by the disc
geometry. The total energy of the alice loop of radius $R$, with a
cheshire charge $Q$ is thus given by:
\begin{equation}
E_{tot} = 2 \pi R {\cal E}_{flux} + \frac{Q^2}{R}~ A
\end{equation}
As mentioned earlier there are two competing, R-dependent, terms and
we may determine the radius of the loop which minimizes the energy,
yielding:
\begin{equation}
R = |Q| \sqrt{\frac{A}{2 \pi {\cal E}_{flux}}}
\end{equation}
Consequently the minimal energy is given by:
\begin{equation}
E = \sqrt{8 \pi {\cal E}_{flux} A} ~|Q|
\end{equation}
Obviously this estimate can only be trusted if the radius of the alice
loop is much larger than the radius of the alice flux and this should
be checked. However this estimate does indicate that the cheshire
charged alice loop can be a stable configuration, where the string
tension is balanced by the repulsion between the magnetic fieldlines.
Interestingly enough this estimate gives an energy proportional to the
magnetic charge, $E\propto |Q|$.

\section{The model}
\label{themodel}
To answer stability questions related to the spherically symmetric
monopole configuration (or the cheshire charged alice ring) we 
consider an explicit model. We use the original tensor alice model
\cite{shankar}, but we could as well have chosen an alternative
\cite{jelper}. We will in fact argue that the results obtained
are quite general and model independent. For completeness and
notational convenience we briefly summarize the model. The action is
given by:
\begin{equation}
S = \int d^4 x\ \left\{\frac{1}{4}F^{a,\mu\nu}F^a_{\mu\nu} + \frac{1}{4}Tr(
D^\mu \Phi D_\mu
\Phi )  - V(\Phi)\right\}\quad,
\end{equation}
where the Higgs field $\Phi = \Phi^{ab}$ is a real, symmetric,
traceless $3\times3$ matrix, i.e., $\Phi$ is in the five dimensional
representation of $SO(3)$ and $D_\mu \Phi =
\partial_\mu\Phi -ie[A_\mu,\Phi]$, with $A_{\mu}=A^a_{\mu}T_a$, where
$T_a$ are the generators of $SO(3)$. The most general renormalizeable
potential is given by
\cite{georgi}:
\begin{equation}
V=-\frac{1}{2}\mu^2Tr\left(\Phi^2\right) -\frac{1}{3}\gamma Tr\left(\Phi^3\right) + \frac{1}{4} \lambda \left(Tr\left(\Phi^2\right)\right)^2\quad,
\label{potential}
\end{equation}
with the parameter $\gamma>0$, since
$(\Phi,\gamma)=(-\Phi,-\gamma)$. For a suitable range of the
parameters in the potential, the gauge symmetry of the model will be
broken to the symmetry of AED. In the ``unitary'' gauge, where the
Higgs field is diagonal, the ground state is (up to permutations)
given by the following matrix:
\begin{equation}
\Phi_0=\left(\begin{array}{ccc}
-f&0&0\\ 0&-f&0\\ 0&0&2f
\end{array}\right)\quad,
\end{equation}
with $f =
\frac{\gamma}{12\lambda}\left(1+\sqrt{1+\frac{24\mu^2\lambda}{\gamma^2}}\right)$.

The full action has four parameters, $e,\mu^2,\gamma,\lambda$, this
number can be reduced to two dimensionless parameters by appropriate
rescalings of the variables.  A physical choice for these
dimensionless parameters is to take the ratio's of the masses that one
finds from perturbing around the homogeneous minimum. To determine
these, we write the action in the unitary gauge where the massless
components of $\Phi$ have been absorbed by the gauge fields. The
physical components of the Higgs field may be expanded as:
\begin{equation}
\Phi(x^\mu)=\Phi_0+\sqrt{2}\phi_1(x^\mu) E_1
+\sqrt{2}\phi_2(x^\mu)~R_3(a(x^\mu))E_2R_3(a(x^\mu))^T\quad,
\end{equation}
with:
\begin{equation}
E_1=\frac{1}{\sqrt{6}}\left(\begin{array}{ccc} -1&0&0\\ 0&-1&0\\ 0&0&2
\end{array}\right)
~;~~E_2=\frac{1}{\sqrt{2}}\left(\begin{array}{ccc} 1&0&0\\ 0&-1&0\\
0&0&0
\end{array}\right)
~;~~E_3=\frac{1}{\sqrt{2}}\left(\begin{array}{ccc} 0&0&1\\ 0&0&0\\
1&0&0
\end{array}\right)
\end{equation}
and $R_i$ are the usual rotation matrices. To second order, the
potential $V(\Phi)$ takes the following form\footnote{It is most
convenient to use $\phi_2$ for the combination $\phi_2e^{ia}$, since
these two Higgs modes, $\phi_2$ and $a$, combine to form one complex
charged field, from now on called $\phi_2$.}:
\begin{equation}
V(\Phi) = const. + (2\mu^2 + \gamma f)\phi_1^2 + 3\gamma f|\phi_2|^2
+~\cdots\quad,
\end{equation}
yielding the two distinct masses of the Higgs modes. Next we look at
the 'kinetic' term, $\frac{1}{4}Tr(D^\mu \Phi D_\mu \Phi )$, of the
Higgs field. Inserting the previous expressions for the Higgs field,
we find:
\begin{equation}
\frac{1}{4}Tr(D^\mu \Phi D_\mu \Phi) = \frac{1}{2}(\partial_\mu\phi_1)^2 +
\frac{1}{2}\left|D_\mu^3 \phi_2\right|^2 +\frac{9}{2}e^2f^2\left[\left(A_\mu^1\right)^2 + \left(A_\mu^2\right)^2\right] +~\cdots\quad,
\end{equation}
with: $D_\mu^3=\partial_\mu - i 2 e A^3_\mu$. The second term shows
that the $\phi_2$ component of the Higgs field carries a charge $2e$
with respect to the unbroken $U(1)$ component $A_\mu^3$ of the gauge
field. The first term describes the usual charge neutral Higgs
particle and the third term yields the mass of the charged gauge
fields. So the relevant lowest order action is given by:
\begin{eqnarray}
S = \int d^4x \left\{ \frac{1}{4} F_{\mu\nu}^a F^{a,\mu\nu} +
\frac{1}{2}(\partial_\mu\phi_1)^2 + \frac{1}{2}\left|D_\mu^3\phi_2\right|^2 -
\frac{1}{2}m_1^2\phi_1^2 \right.\\ \nonumber
\left.- \frac{1}{2}m_2^2|\phi_2|^2 -
\frac{1}{2}m_A^2\left[\left(A_\mu^1\right)^2+\left(A_\mu^2\right)^2\right]+~\cdots\right\}\quad,
\end{eqnarray}
with $m_1^2=4\mu^2 + 2\gamma f$, $m_2^2=6\gamma f$ and
$m_A^2=9e^2f^2$.

Two degrees of freedom of the five dimensional Higgs field are 'eaten'
by the broken gauge fields, one degree of freedom forms the real
neutral scalar field and two degrees of freedom form the complex
(doubly charged) scalar field. To specify a point in the classical
parameter space we may, up to irrelevant rescalings, use the
dimensionless mass ratio's $\frac{m_1}{m_2}$ and $\frac{m_A}{m_2}$. We
note that the value of $\frac{m_1}{m_2}$ needs to be larger or equal
to $\frac{1}{3}$, because for smaller values of $\frac{1}{3}$ the
groundstate corresponds to the symmetric unbroken vacuum. Note that we
found three mass scales in this problem. For the simulations, which we
will describe in the next section, this means that we will we have to
deal with three different length scales.

\section{Numerical simulations}
In this section we will describe in some detail the numerical
simulations we performed to determine the instability and the
meta-stability regions for the spherically symmetric monopole,
in the parameter space of the model. First though, we introduce the
ansatz. We end the section with the discussion of a typical set of
numerical experiments.

\subsection{The variational ansatz}
As mentioned before we will use a variational approach. In such an
approach the configurations one finds are typically not exact
solutions to the equations of motion. However as the ansatz we will
use contains the ansatz for the exact spherically symmetric solution
we may still study the instability and the meta-stability of this
solution. This means that the instability and the meta-stability
regions we will find for the monopole, are lower bounds, in the sense
that those instability regions can only become larger as the ansatz
becomes less restrictive.

As we expect the competing configuration of the spherically symmetric
magnetic monopole to be the magnetically cheshire charged alice ring,
we base our ansatz on cylindrical symmetry, also see
\cite{jelper2}. The ansatz we will use also has a reflection symmetry
with respect to the $z$$=$$0$-plane. We impose this reflection
symmetry to eliminate the (almost) zero mode in the energy due to the
position of the defect along the $z$-axis. The ansatz for the Higgs
field is:
\begin{equation}
\Phi(z,\rho,\theta=0)=\phi_1(z,\rho) E_1+ \phi_2(z,\rho) E_2 + \phi_3(z,\rho) E_3
\end{equation}
and
\begin{equation}
\Phi(z,\rho,\theta)=R_3(\theta)\Phi(z,\rho,\theta=0)R_3(\theta)^T\quad.
\end{equation}
The ansatz for the gauge fields is simply given by
$eA_i^j=-\epsilon_{ijk}\frac{x^k}{x^2}A(z,\rho)$, very similar to the
one for the spherically symmetric monopole \cite{'thooft}, except that
we allow $A(z,\rho)$ to depend on $\rho$ and $z$ and not only on
$r=\sqrt{\rho^2+z^2}$. The boundary conditions for $r\to\infty$ are
the boundary conditions of the spherically symmetric monopole
\cite{shankar}, i.e., $A(z,\rho)$ goes to one and the Higgs field to
$\Phi(z,\rho,\theta)=R_3(\theta)R_2\left(\arccos\left(\frac{z}{r}\right)\right)
\Phi_0 R_2\left(\arccos\left(\frac{z}{r}\right)\right)^TR_3(\theta)^T$. The boundary
conditions at $\rho=0$ and $z=0$ follow by imposing the cylindrical
and reflection symmetry and are given in the table below:
\begin{displaymath}
\begin{array}{|c||c|c|}\hline
&\rho=0&z=0\\\hline
\phi_1&\partial_\rho\phi_1=0&\partial_z\phi_1=0\\
\phi_2&\partial_\rho\phi_2=\phi_2=0&\partial_z\phi_2=0\\
\phi_3&\phi_3=0&\phi_3=0\\
A&\partial_\rho A=0&\partial_zA=0\\\hline
\end{array}
\end{displaymath}
It is easy to see that these boundary conditions are also met by the
spherically symmetric monopole, so it is indeed contained in our more
general ansatz. With the help of this model and ansatz we study the
stability of the spherically symmetric magnetic monopole. Before we
present the results we describe the numerical methods we employ, and a
typical set of experiments.

\subsection{Some numerical details}
\begin{figure}[!htb]
\begin{center}
\makebox[7.9cm]{\psfig{figure=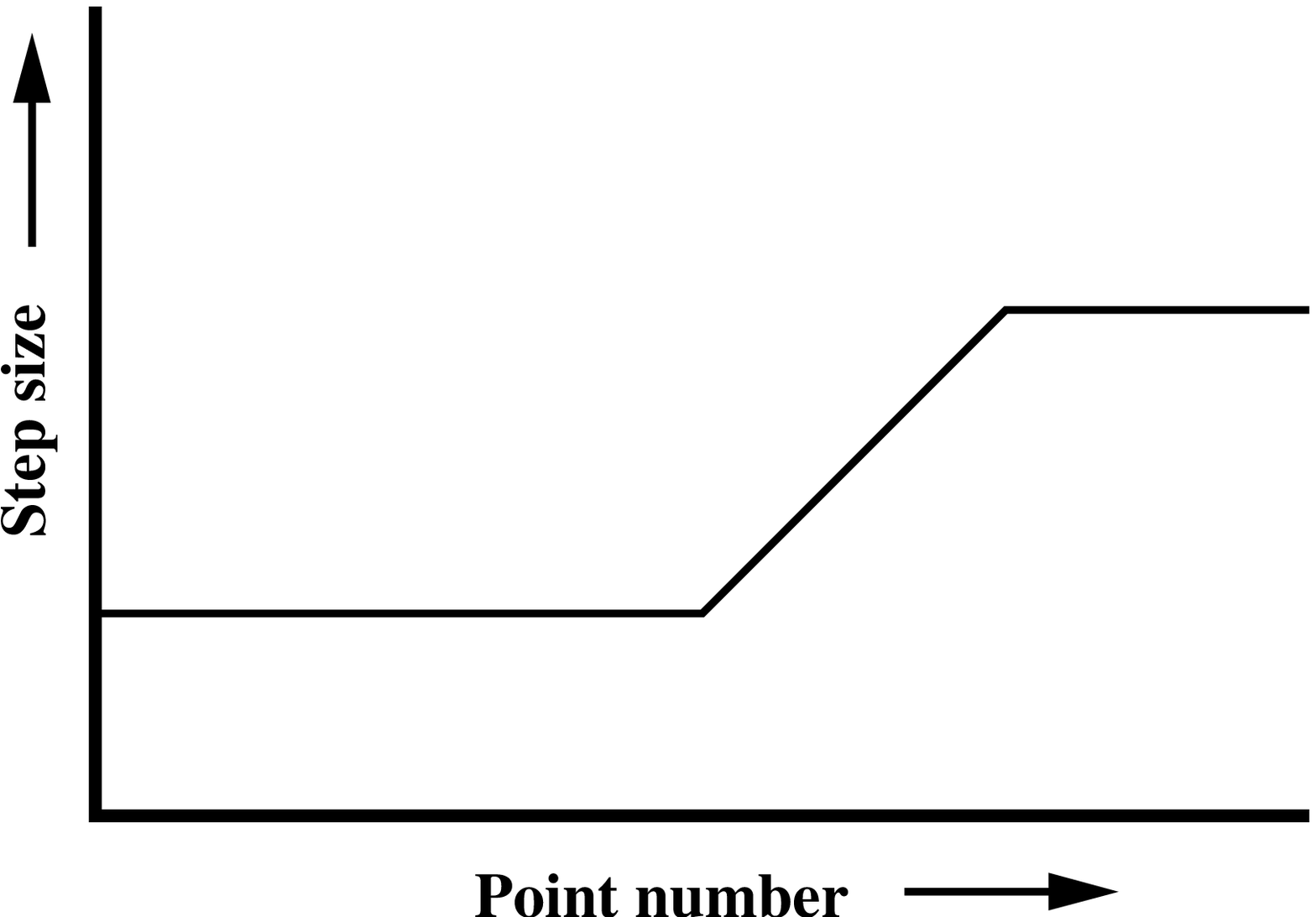,width=7cm,angle=0}}
\makebox[7.9cm]{\psfig{figure=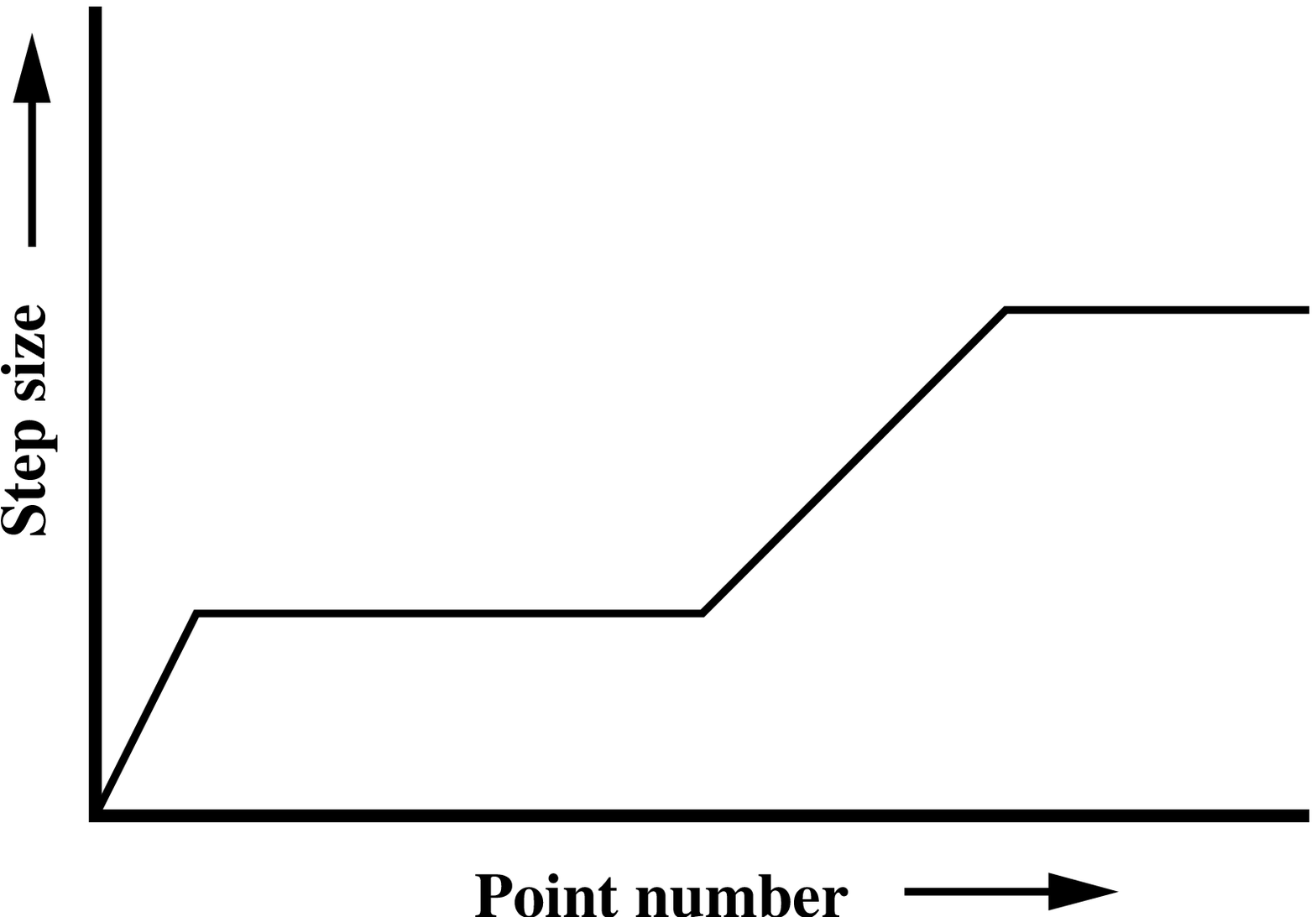,width=7cm,angle=0}}
\makebox[7.9cm][c]{\footnotesize{(a)}}
\makebox[7.9cm][c]{\footnotesize{(b)}}
\caption[somethingelse]{\footnotesize The two different types of
lattices. The first part covers the core structure of the defects
where the second part captures the region where the gauge fields -
which have a much smaller mass - show nontrivial behavior. The only
difference of a lattice of type (b) with respect to lattice of type
(a) is a few extra lattice points near $r=0$ and $z=0$.}
\label{lattice}
\end{center}
\end{figure}

We mentioned in section \ref{themodel} that the AED model has three
mass scales, i.e., three relevant length scales. However we examine a
region of parameter space in which only two of those are
relevant. That is the core geometry of the defect and the region where
the Higgs field is not in the vacuum manifold on the one hand, and the
inverse mass of the gauge fields - which is typically much larger - on
the other. To be able to adequately capture both scales we use a space
and configuration dependent lattice spacing. In fact we will use two
types of lattices. In figure \ref{lattice} we schematically give the
step sizes as a function of the point number (in one dimension). The
two dimensional lattices we will use have the same type of lattice in
both directions.

The only difference between the two lattices is that in the second,
figure \ref{lattice}(b), there are extra lattice points near
$r=0$ and $z=0$ and thus it has a more lattice points. However the
same part of space is covered by both lattices. Although the
difference between the two lattices is quite small we will encounter a
specific lattice dependence, which turns out it be an artifact,
but nevertheless will be of use later on.

To obtain the minima of the energy within the ansatz for the different
values of the parameters of the model, we used a Monte Carlo based
cooling method. In this method one introduces a temperature and gives
a configuration with energy $E$ a weight factor equal to
$e^{-\frac{E}{T}}$. In the limit of $T\to 0$ only the configuration
with the lowest energy survives. Assuming there are no flat directions
this procedure selects an unique configuration. With the help of a
Monte Carlo mechanism different configurations are sampled. We keep
sampling at a specific temperature as long as the energy of the system
averaged over a preset number of sweeps\footnote{In one sweep all
variables undergo one Monte Carlo step, i.e., are allowed to change
once.} (typically 10) decreases and do this a minimal number of times
(typically 3). When this average energy no longer decreases we lower
the temperature (typically by 10\%) and repeat the process until the
total energy drop at a specific temperature becomes lower then a
predetermined relative energy change (typically
$\sim10^{-6}$\%). During this process we keep the acceptance rate of
the Monte Carlo steps locally fixed for all fields. This means that we
introduced a maximum step size for each field at each position and
automatically adjust it to get the preferred acceptance rate. Since
the change due to temperature change is easy captured we determined
the desired maximum step sizes in the beginning of the cooling
mechanism and further only correct for the trivial temperature
changes, i.e., as $T$ changes in $a T$, $\delta_{field}$ changes in $
\sqrt{a}\delta_{field}$. This assumes that the energy change depends
quadratically on the field change which is what one naively expects
near a stable configuration, and indeed, that criterion worked nicely.

The procedure we just described is used to determine stable
configurations within our ansatz for the different values of the
parameters of the model. To be able to determine the meta-stability
and instability regions of the monopole in the parameter space of the
model we perform hysteresis type of experiments. This means that we
determine the lowest energy configuration for a specific point in the
parameter space and use this configuration as the starting point for
the determination of a stable configuration at a point nearby in the
parameter space. As the change in the parameter space is only small
one would expect the new configuration also to be close to the
previous one. We determine the new configuration again with the help
of the cooling method\footnote{The temperature at which the secondary
cooling process starts is smaller than the starting temperature of the
initial cooling process. This starting temperature of the secondary
cooling processes determines the energy barriers which can be
overcome. Only in the limit where this temperature goes to zero can
one really claim that a configuration becomes unstable. However for
finite temperature one can still show the instability under small, but
finite, perturbations. This is what we mean if we claim to find an
instability of a specific type of configuration.}. We repeat this
process and move along a trajectory in the parameter space and back
again. We choose these paths such that the lowest energy
configurations on each end of the trajectory are different type of
configurations: a spherical monopole and a cheshire charged alice
ring. To keep things numerically simple we keep the mass of the gauge
fields constant during a hysteresis experiment. This restriction
selects specific trajectories in the two dimensional parameter space,
which are given by:
\begin{equation}
\frac{m_A}{m_2}=\frac{2~\left(\frac{m_A}{m_2}\right)_\frac{1}{3}}{3\sqrt{\frac{1}{3}+\left(\frac{m_1}{m_2}\right)^2}}\quad, 
\end{equation}
with $\left(\frac{m_A}{m_2}\right)_\frac{1}{3}$ the value of
$\frac{m_A}{m_2}$ at the minimal value of
$\frac{m_1}{m_2}=\frac{1}{3}$.\\ This restriction allows us to keep
the total space covered by the same lattice throughout the
hysteresis. In a single hysteresis experiment the size of the core
does not change much, so we keep the lattice fixed throughout a single
hysteresis experiment.
\begin{figure}[!htb]
\begin{center}
\makebox[7.9cm]{\psfig{figure=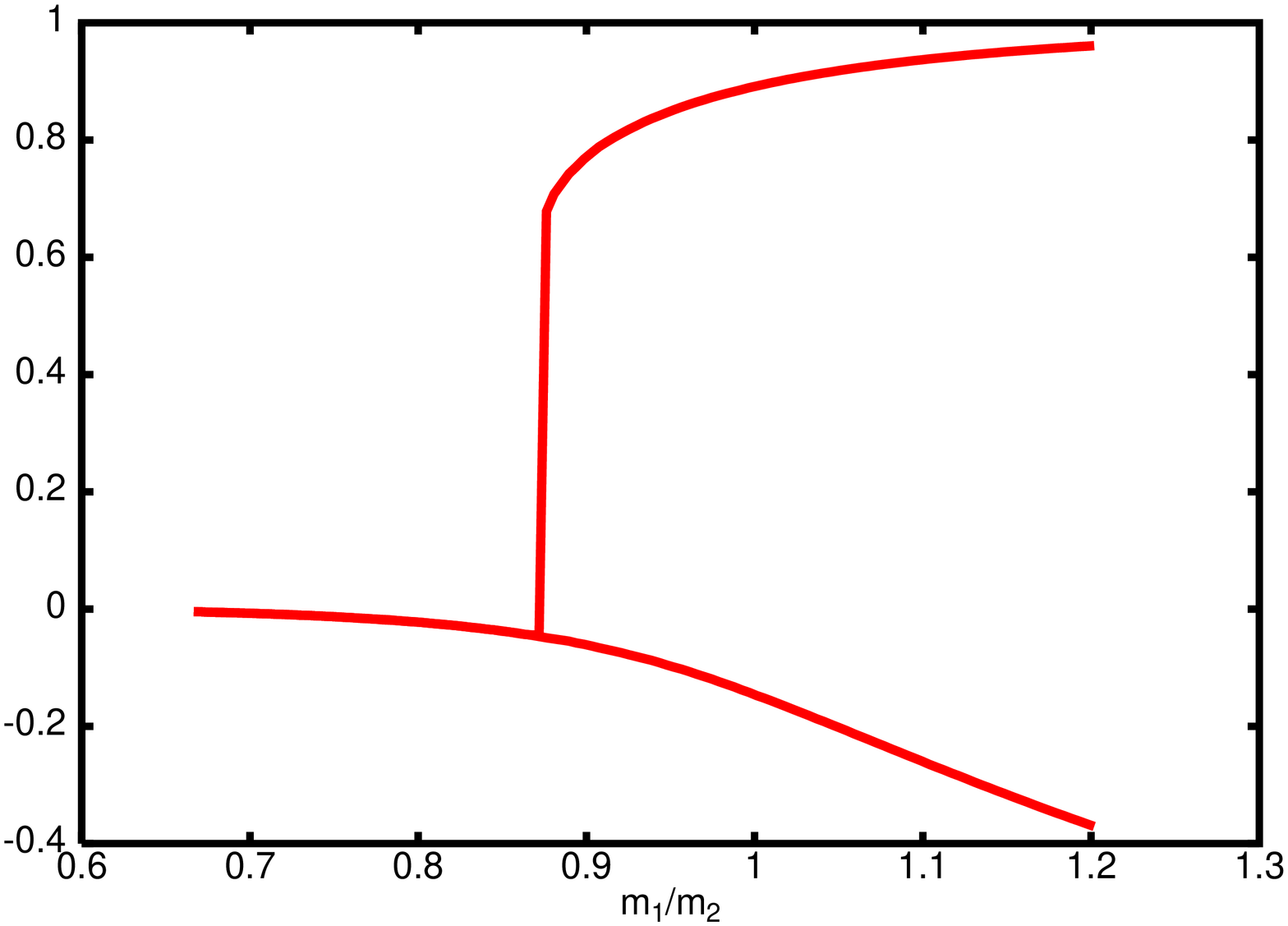,width=7.9cm,angle=270}}
\makebox[7.9cm]{\psfig{figure=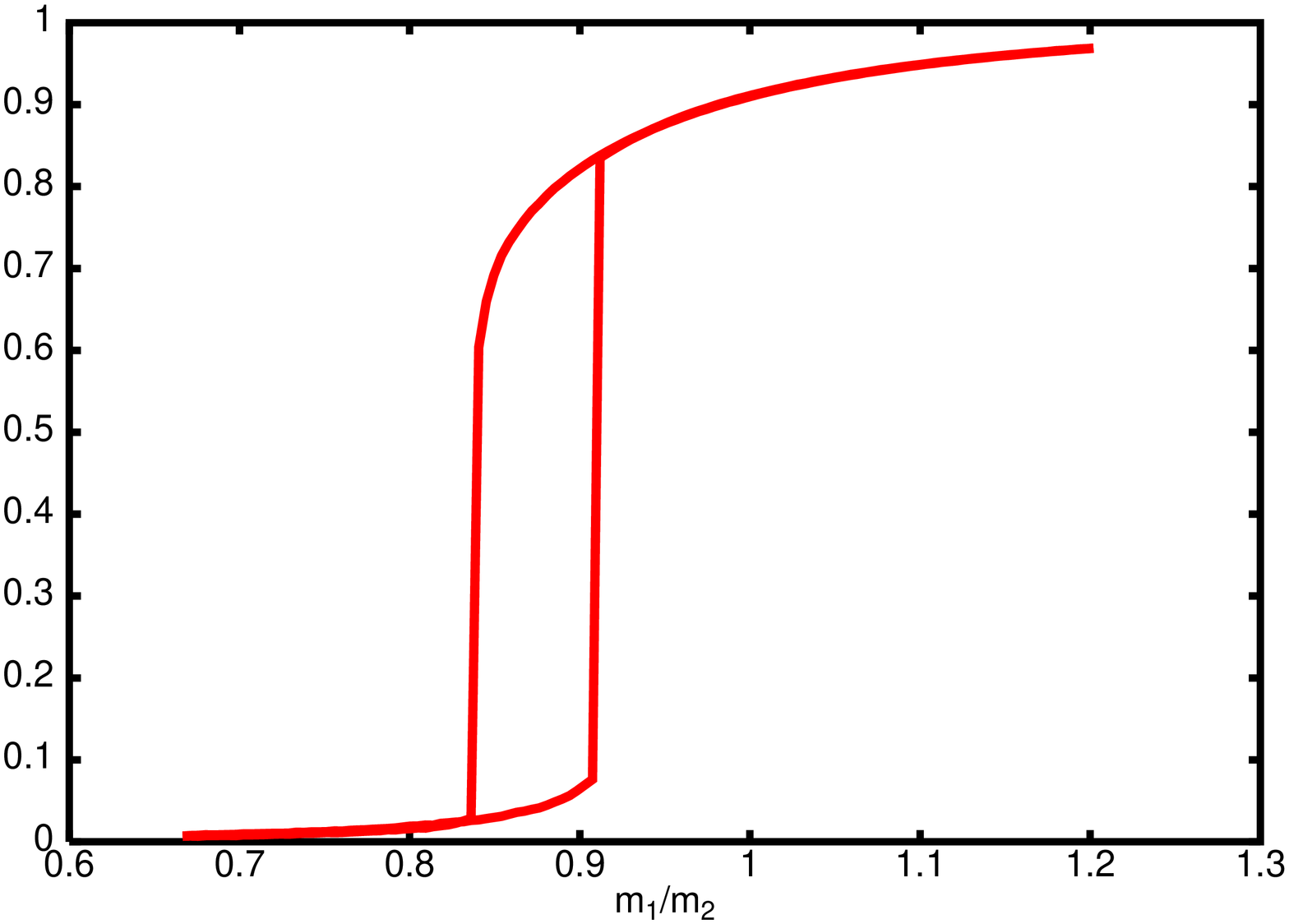,width=7.9cm,angle=270}}
\makebox[7.9cm][c]{\footnotesize{(a)}}
\makebox[7.9cm][c]{\footnotesize{(b)}}
\caption[somethingelse]{\footnotesize The two typical results of the
hysteresis experiments for the different lattice types projected on
the $\frac{m_1}{m_2}$-axis. Figure (a) corresponds to a lattice of
type (a) and figure (b) corresponds to a lattice of type (b).}
\label{hysteresus}
\end{center}
\end{figure}

\begin{figure}[!htb]
\begin{center}
\makebox[7.9cm]{\psfig{figure=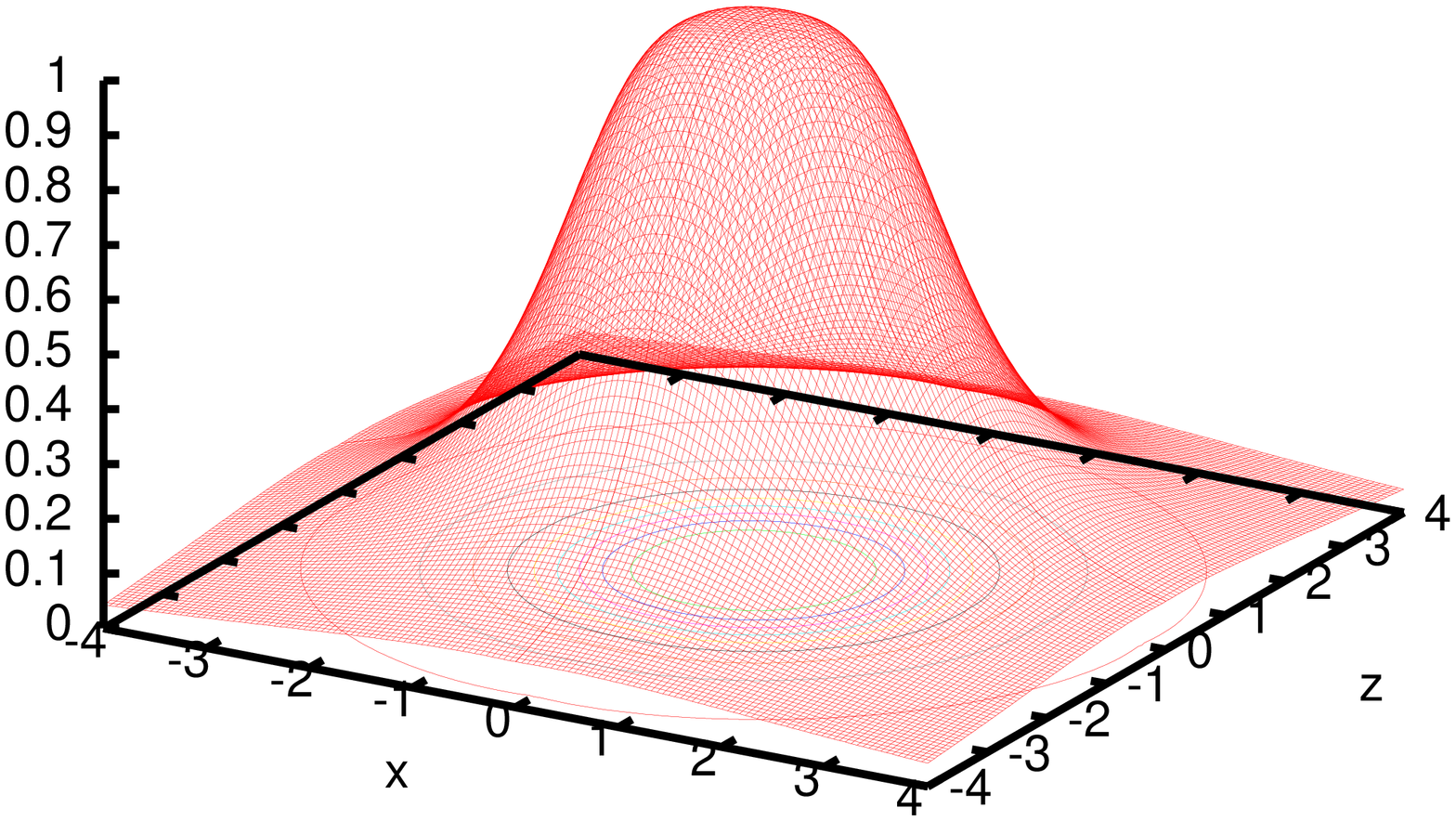,width=8.0cm,angle=270}}
\makebox[7.9cm]{\psfig{figure=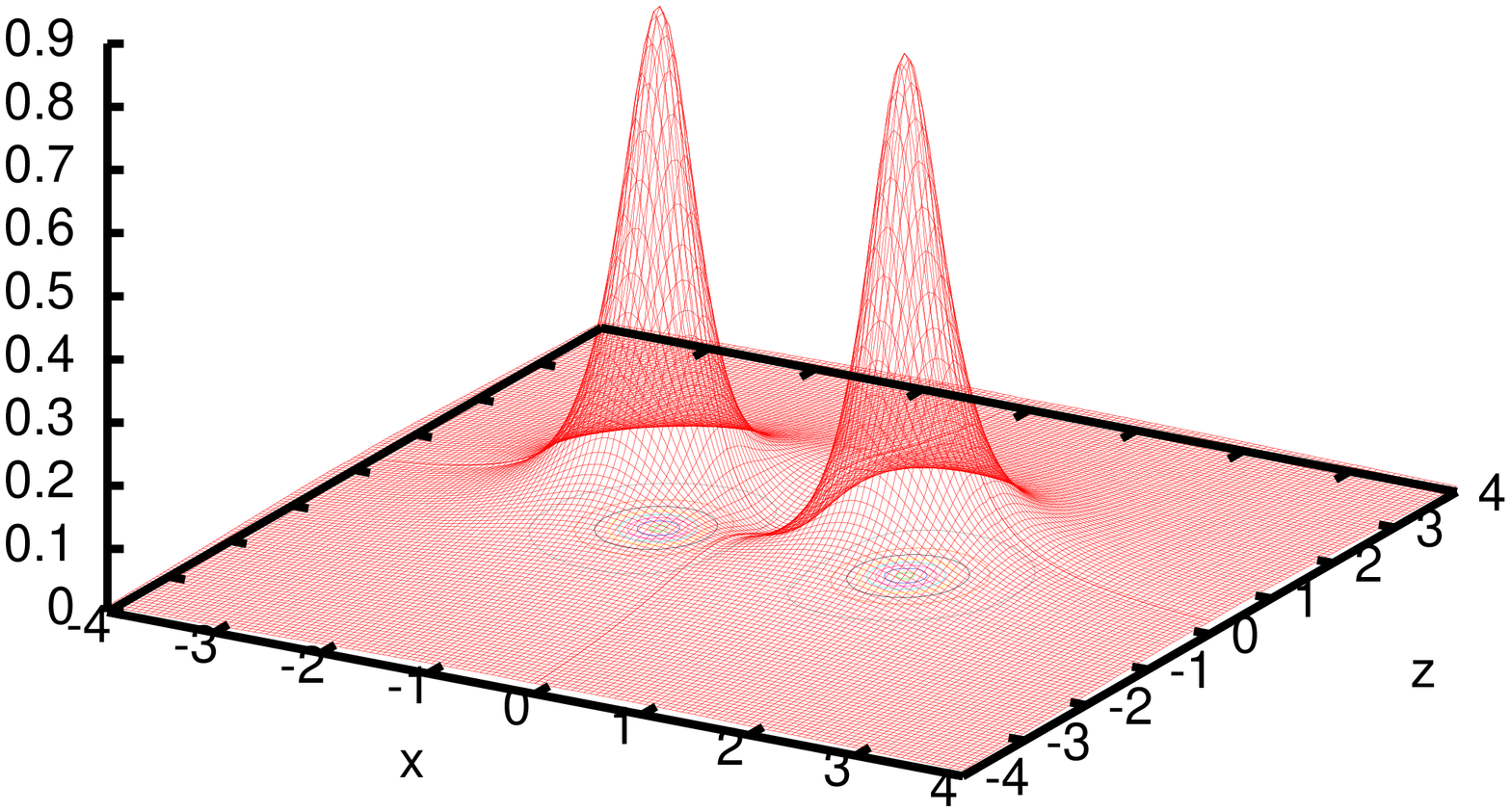,width=8.0cm,angle=270}}
\makebox[7.9cm][c]{\footnotesize{(a)}}
\makebox[7.9cm][c]{\footnotesize{(b)}}
\makebox[15.8cm]{\psfig{figure=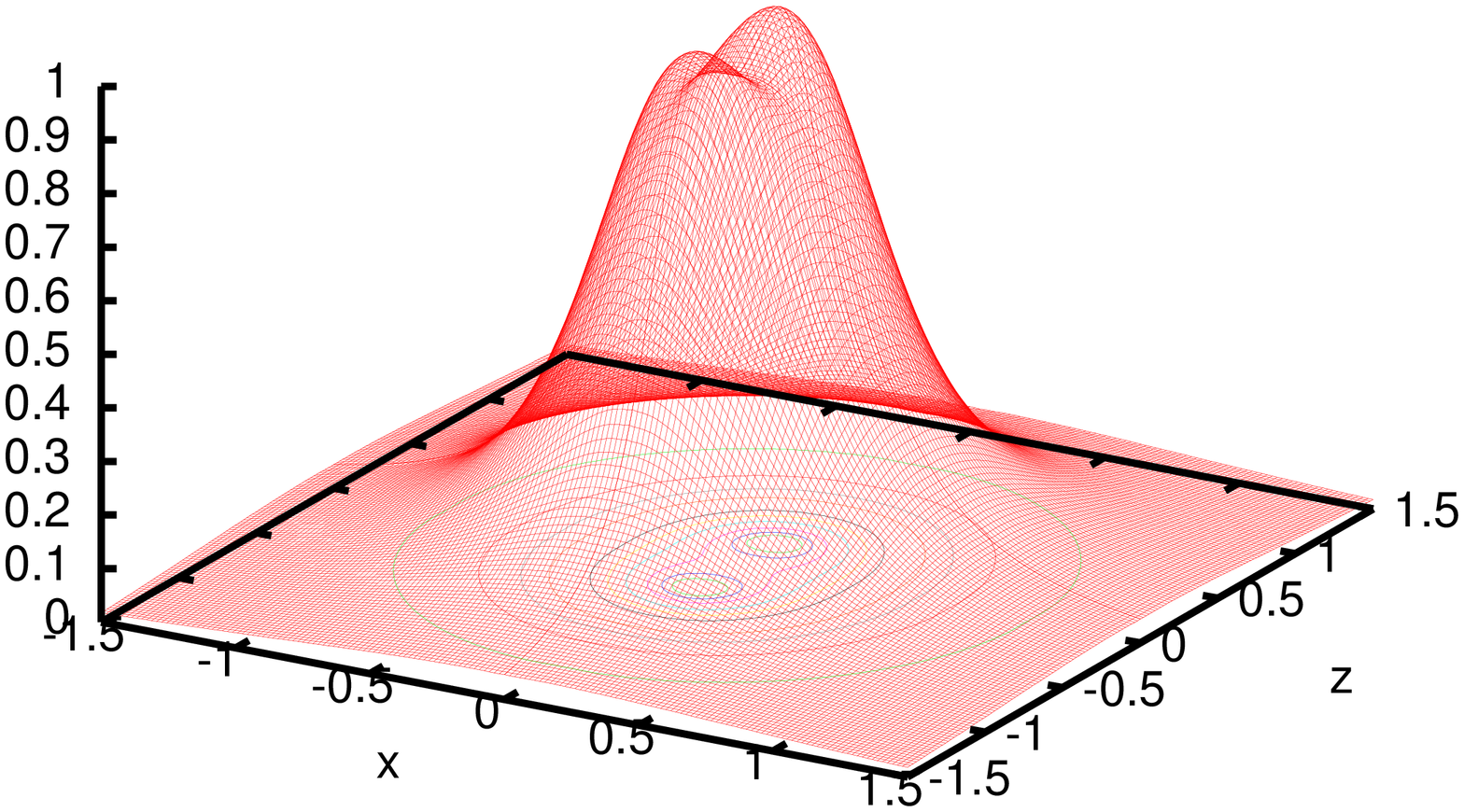,width=8.0cm,angle=270}}
\makebox[15.8cm][c]{\footnotesize{(c)}}
\caption[somethingelse]{\footnotesize These figures show the three
typical core structures we encountered in the numerical
simulations. All three figures show a slice of the core structure at
$y=0$ an in the figures the value of $1-\frac{Tr\Phi^2}{6f^2}$ is
plotted. In figure (a) we plotted the core structure of the
spherically symmetric magnetic monopole at $\frac{m_1}{m_2}=0.57$ and
$\frac{m_A}{m_2}=0.0095$. In figure (b) we plotted the core structure
of the magnetically cheshire charged alice ring at
$\frac{m_1}{m_2}=0.88$ and $\frac{m_A}{m_2}=0.0073$ and in figure (c)
we plotted the core structure of the split core configuration at
$\frac{m_1}{m_2}=1.12$ and $\frac{m_A}{m_2}=0.0425$.}
\label{defects}
\end{center}
\end{figure}

Figures \ref{hysteresus}(a) and (b) show the typical hysteresis
results. The figures show the values of the field variable
$\frac{\phi_1}{\sqrt{6}f}$ at $r=z=0$. This variable is a normalized
order parameter in the sense that its value equals zero for the
spherical monopole, and one for the alice ring. A negative value of
the order parameter is also possible and corresponds to a new type of
configuration, the so called {\it split core}, see figure
\ref{defects}(c) and also \cite{gartland2,gartland}. Figure
\ref{hysteresus}(b) shows what one would expect for a hysteresis type
of experiment for a lattice of type (b) while figure
\ref{hysteresus}(a) shows a different behavior and corresponds to a
lattice of type (a).

In figure \ref{hysteresus}(a) the split core configuration appears and
although this configuration typically has more energy then a cheshire
charged alice ring configuration, it does appear to be meta-stable.
In \cite{gartland2,gartland} this object was discussed for the global
analog of AED, a nematic liquid crystal theory, and it was argued that
this configuration is due to the cylindrical symmetry restriction of
the ansatz used to explore the solution-space. Here we found that it
might as well be just a lattice artifact as it depends on the lattice
type one uses in the hysteresis experiments. Although this splitcore
configuration may be viewed as an undesirable feature caused by
lattice and/or symmetry artifacts, we will see that it can be turned
into a useful tool.

\subsection{A typical set of experiments}
\label{experiments}
\begin{figure}[!htb]
\begin{center}
\makebox[7.9cm]{\psfig{figure=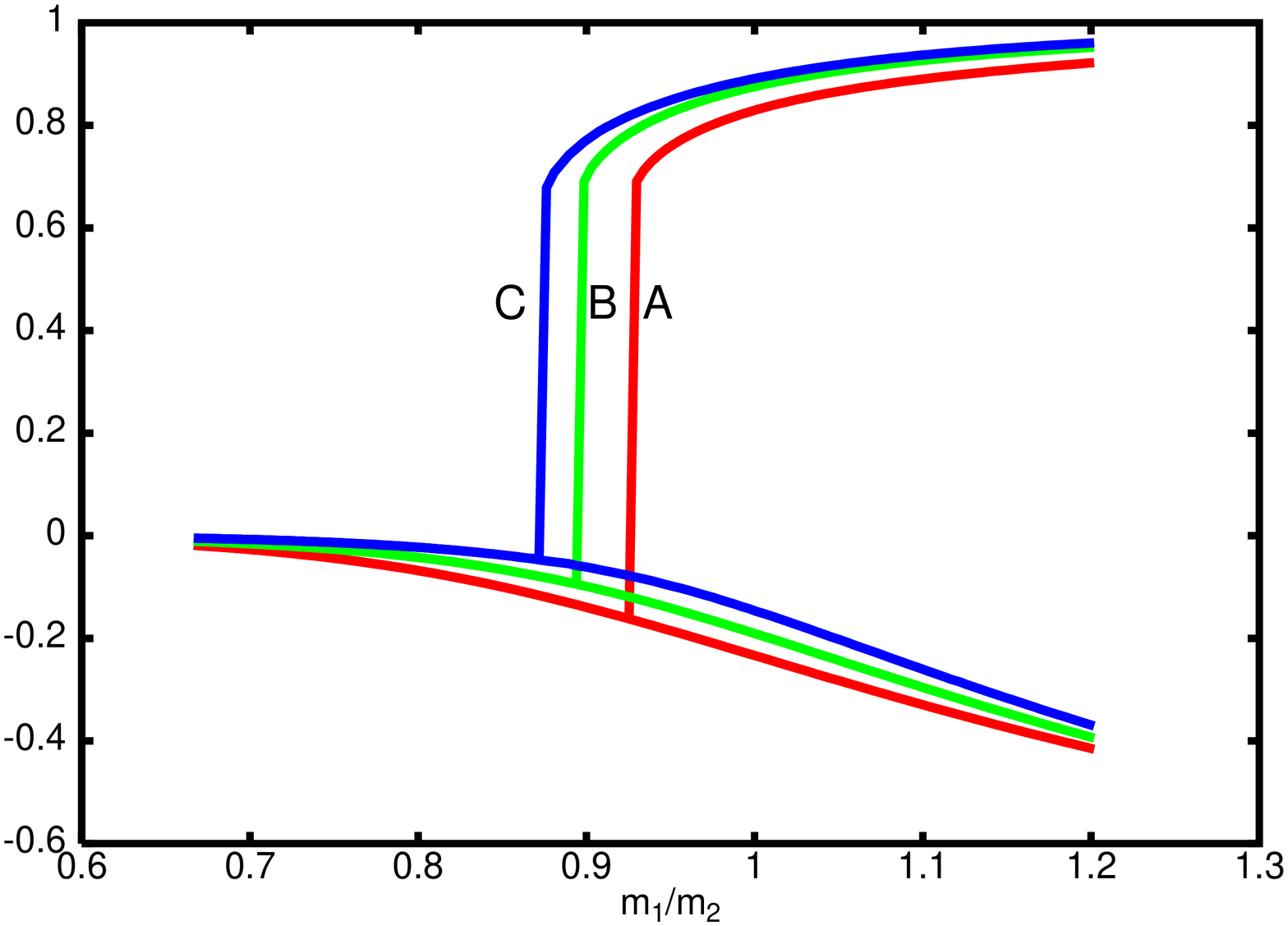,width=7.9cm,angle=270}}
\makebox[7.9cm]{\psfig{figure=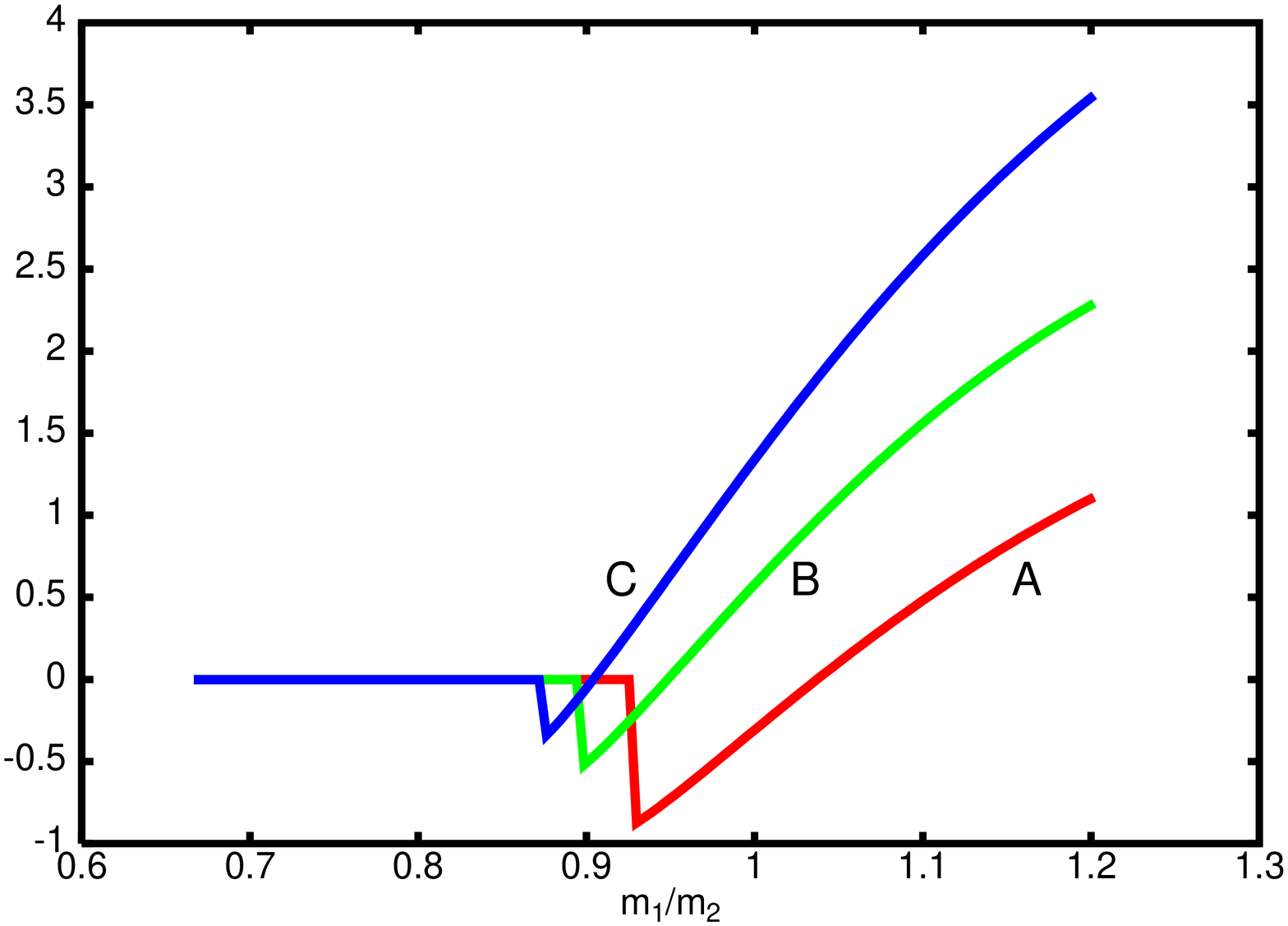,width=7.9cm,angle=270}}
\makebox[7.9cm][c]{\footnotesize{(a)}}
\makebox[7.9cm][c]{\footnotesize{(b)}}
\makebox[7.9cm]{\psfig{figure=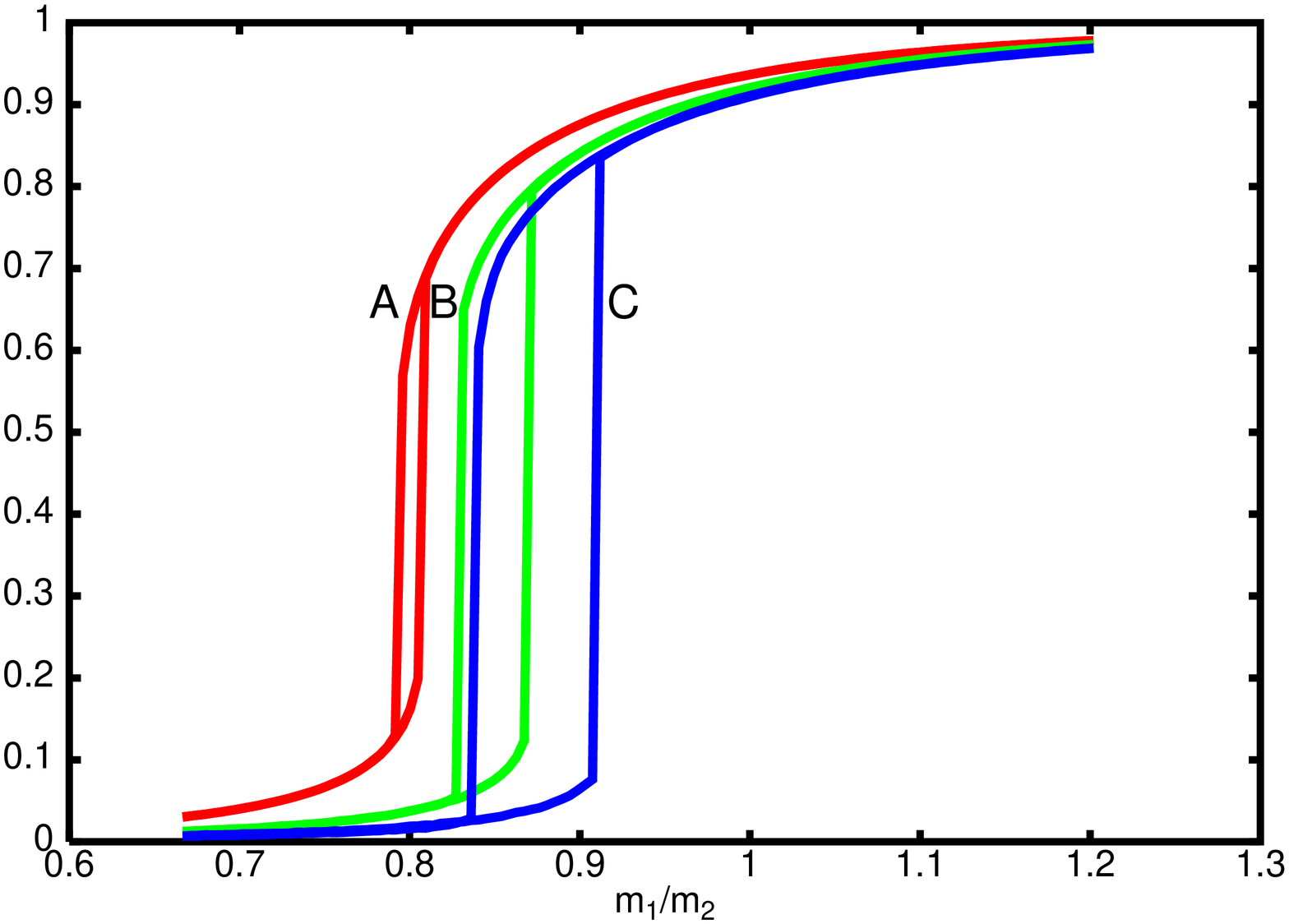,width=7.9cm,angle=270}}
\makebox[7.9cm]{\psfig{figure=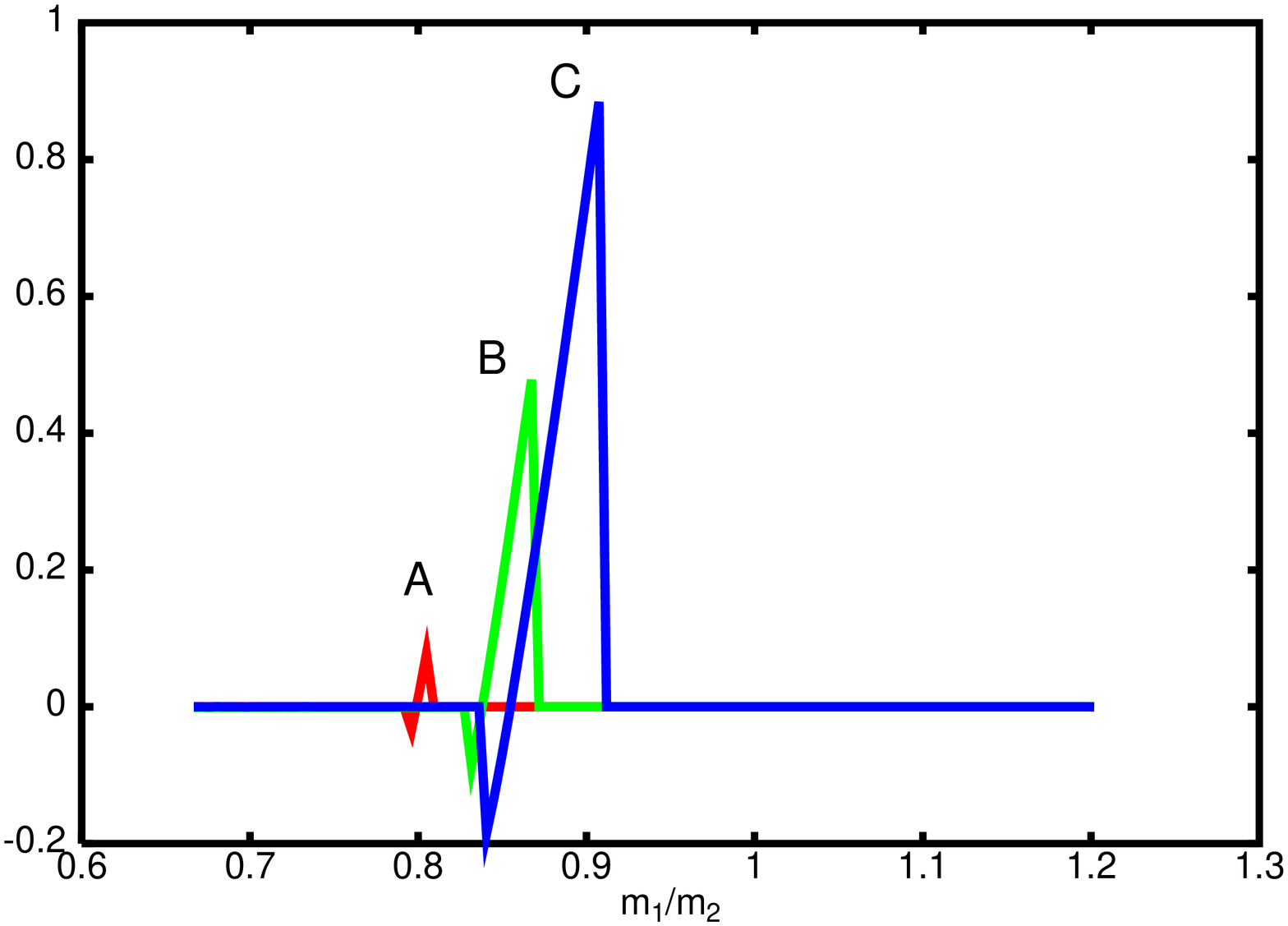,width=7.9cm,angle=270}}
\makebox[7.9cm][c]{\footnotesize{(c)}}
\makebox[7.9cm][c]{\footnotesize{(d)}}
\caption[somethingelse]{\footnotesize These figures show the results
of a typical set of experiments of a specific line in the parameter
space, with the results projected on the $\frac{m_1}{m_2}$-axis.
Figures (a) and (b) correspond to a lattice of type (a), while figures
(c) and (d) correspond to a lattice of type (b). In the figures (a)
and (c) the value of $\frac{\phi_1(0,0)}{\sqrt{6}f}$ is plotted, while
in the figures (b) and (d) the relative energy differences of the two
branches of the hysteresis plotted in pro-mils. The lines A, B and C
correspond to 25$\times$25 (27$\times$27), 50$\times$50 (54$\times$54)
and 100$\times$100 (108$\times$108) lattice points respectively
describing the core structure.}
\label{typical}
\end{center}
\end{figure}

As mentioned before we do hysteresis type of experiments along
specific trajectories in the parameter space of the model. Next we
look at a typical set of such experiments. We did the hysteresis
experiment for three different numbers of lattice points, where we
only changed the number of lattice points describing the core of the
configuration. One could of course also change the number of points
outside the core but we found that that did not make any difference in
the observables we examined and did not affect the stability of the
configurations. We performed numerical simulations with 25$\times$25
(27$\times$27), 50$\times$50 (54$\times$54) and 100$\times$100
(108$\times$108) lattice points describing the core structure for each
line in the parameter space we considered. The figures
\ref{typical}(a-d) are the typical results of such an
investigation. The figures show two different observables: the
relative energy difference of the two branches of the hysteresis, and
the quantity $\frac{\phi_1(0,0)}{\sqrt{6}f}$. The figures
\ref{typical}(a) and \ref{typical}(b) belong to lattices of type (a)
and the figures \ref{typical}(c) and \ref{typical}(d) belong to
lattices of type (b). All figures show the results obtained with the
three different number of lattices points, with the lines A, B an C
corresponding to 25$\times$25 (27$\times$27), 50$\times$50
(54$\times$54) and 100$\times$100 (108$\times$108) points describing
the core structure respectively.

Let us first compare the figures \ref{typical}(a) and
\ref{typical}(c). These figures show the values of
$\frac{\phi_1(0,0)}{\sqrt{6}f}$. In both figures we see that
qualitatively the lines A, B and C do not differ very much, but
quantitatively they do. The steep part of the lines corresponds to the
local instability of the cheshire charged alice ring and the spherical
monopole. In both figures \ref{typical}(a) and \ref{typical}(c) we see
that the alice ring becomes locally unstable with respect to the
monopole for low enough values of $\frac{m_1}{m_2}$. In figure
\ref{typical}(c) we see that for large enough values of
$\frac{m_1}{m_2}$ the monopole becomes unstable with respect to the
alice ring, whereas in figure \ref{typical}(a) we see that the
monopole slowly transforms into a split core configuration, see figure
\ref{defects}(c), and does not become locally unstable with respect to
the alice ring. In figure \ref{typical}(b) we do see that this
splitcore configuration is globally unstable with respect to the alice
ring configuration as it costs more energy.

Now let us examine and compare the figures \ref{typical}(b) and
\ref{typical}(d). First we note that the relative energy differences
are very small, being of the order of pro-mils. This is one of the
reasons why the minimal relative energy difference step in the cooling
mechanism is chosen so small, of the order of $10^{-5}$ pro-mil. In
the regions where both branches of the hysteresis give the same
configuration, see figures \ref{typical}(a) and \ref{typical}(c), the
relative energy difference is equal to zero on the scale of the
figures \ref{typical}(b) and \ref{typical}(d). The other segment of
the curves is the interesting part, in figure \ref{typical}(b) and
\ref{typical}(d). The point where this segment of the curves goes to zero
is the point where the spherically magnetic monopole solution is no
longer the lowest energy configuration within our ansatz, i.e., it is
the point where the monopole becomes globally unstable. As we use more
lattice points to probe this meta-stability this point moves only very
slightly. Typically one should extrapolate the results to infinitely
many points, but here we can use a different approach. We will exploit
the results of the two different types of lattice. In the limit of
infinitely many lattice points both lattice types will move to the
same point. However in the figure \ref{typical}(b) we see that this
point in approached from the right while in figure \ref{typical}(d) it
is approached from the left\footnote{We observed this feature for all
the trajectories through parameter space we considered.} when
increasing the number of lattice points. Obviously this helps us to
determine the position of the point where the monopole becomes
globally unstable, as well as the error in the position of the
point. So we may turn the lattice dependence into a useful tool to
determine the global stability of the spherically symmetric monopole
solution.

For most trajectories through the parameter space the same feature can
be used to determine the point where the alice ring configuration
becomes locally unstable with respect to the monopole. However this
point is not as interesting, since the alice ring configuration is not
necessarily an exact solution to the equations of motion. Both lattice
types also show a local instability of the spherically symmetric
monopole solution. In figure \ref{typical}(c) this happens at a clear
point, but from figure \ref{typical}(a) where the monopole changes
into a splitcore configuration, it is a bit harder to fix the point
where this happens as it appears to be a continuous process. Although
the position of this point is unclear, from figure \ref{typical}(a),
it is at least clear that this point moves in the same direction for
both types of lattices as follows from figures \ref{typical}(a) and
\ref{typical}(c).

With the help of the results of both types of lattice we determine the
global instability point of the spherically symmetric monopole
solution and local instability point of alice ring configuration. To
determine the monopole instability point we only use the results from
the lattices of type (b) as they clearly show the point where the
monopole solution becomes locally unstable.

\section{The meta-/instability results}
\label{someresults}
Let us now turn to results of our investigations. First we describe
how we extracted the results from the hysteresis type of
experiments. There are two important results. In the first place,
there is the line bounding the region in parameter space where the
spherically symmetric magnetic monopole solution is no longer the
lowest energy configuration. Crossing that line the solution only
becomes meta-stable. Although our variational method does not prove
that the configuration which has the lowest energy is a cheshire
charged alice ring solution, in that case it actually does imply it,
as we find that a cheshire charged alice ring configuration exists for
those parameter values. In the second place there is the other line
bounding the region where the spherically symmetric magnetic monopole
solution is no longer a locally stable solution. Finally we also
determined the line at which the alice ring configuration becomes locally
unstable, but as explained before, this line is of least interest as
the alice ring configuration (with our ansatz) is typically not a
solution to the equations of motion of the model.

\begin{figure}[!htb]
\begin{center}
\makebox[7.9cm]{\psfig{figure=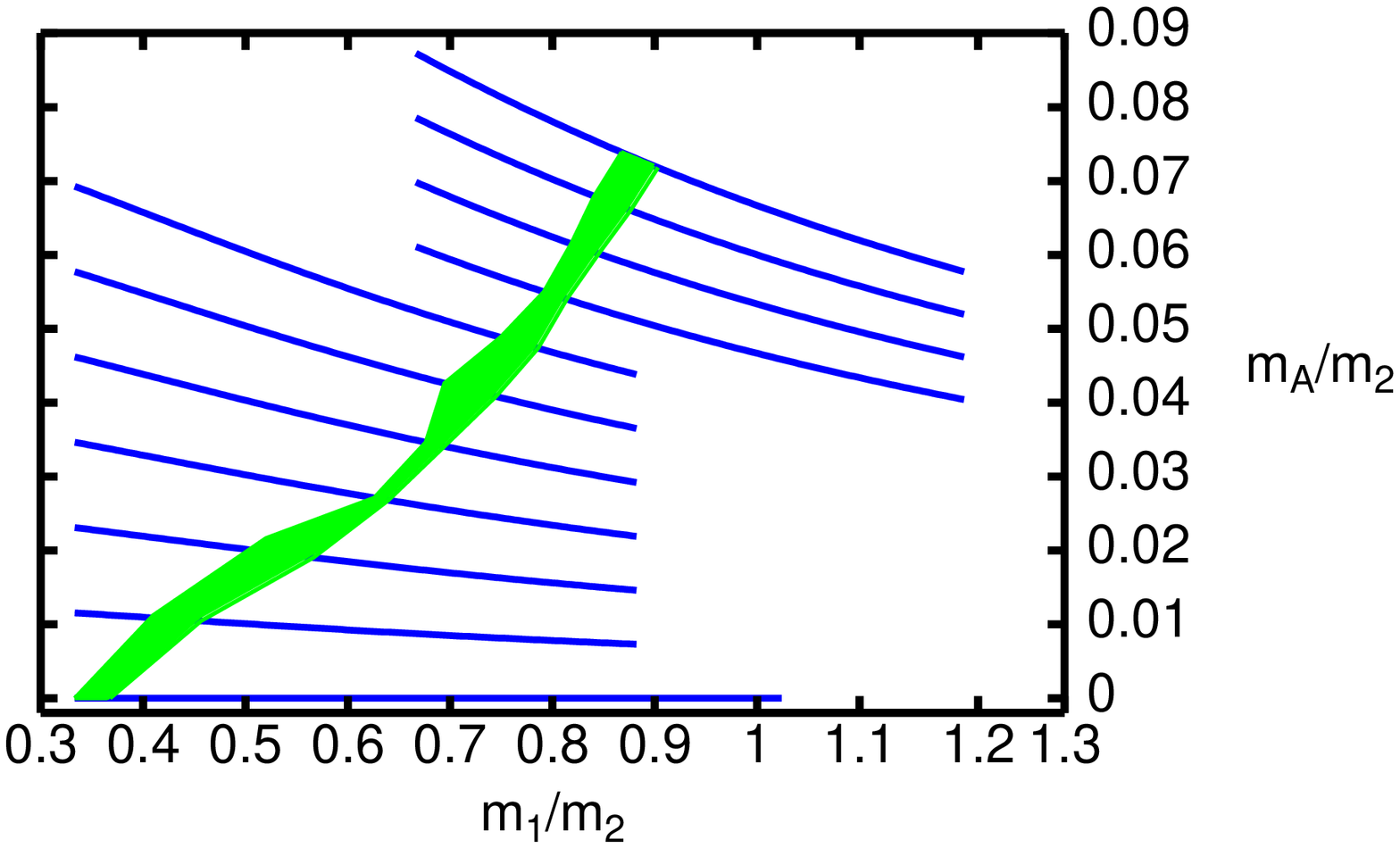,width=7.9cm,angle=270}}
\makebox[7.9cm]{\psfig{figure=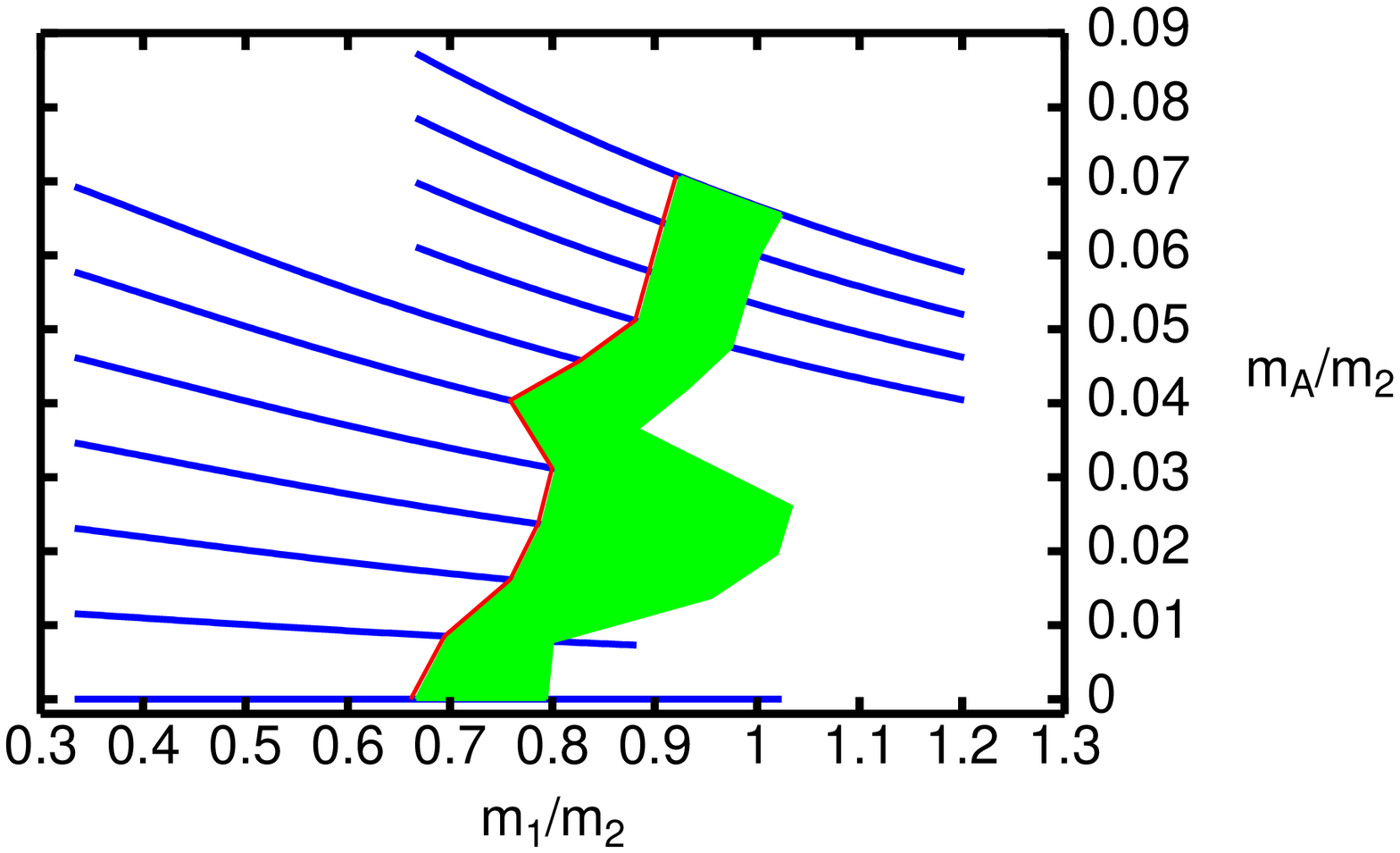,width=7.9cm,angle=270}}
\makebox[7.9cm][c]{\footnotesize{(a)}}
\makebox[7.9cm][c]{\footnotesize{(b)}}
\makebox[15.8cm]{\psfig{figure=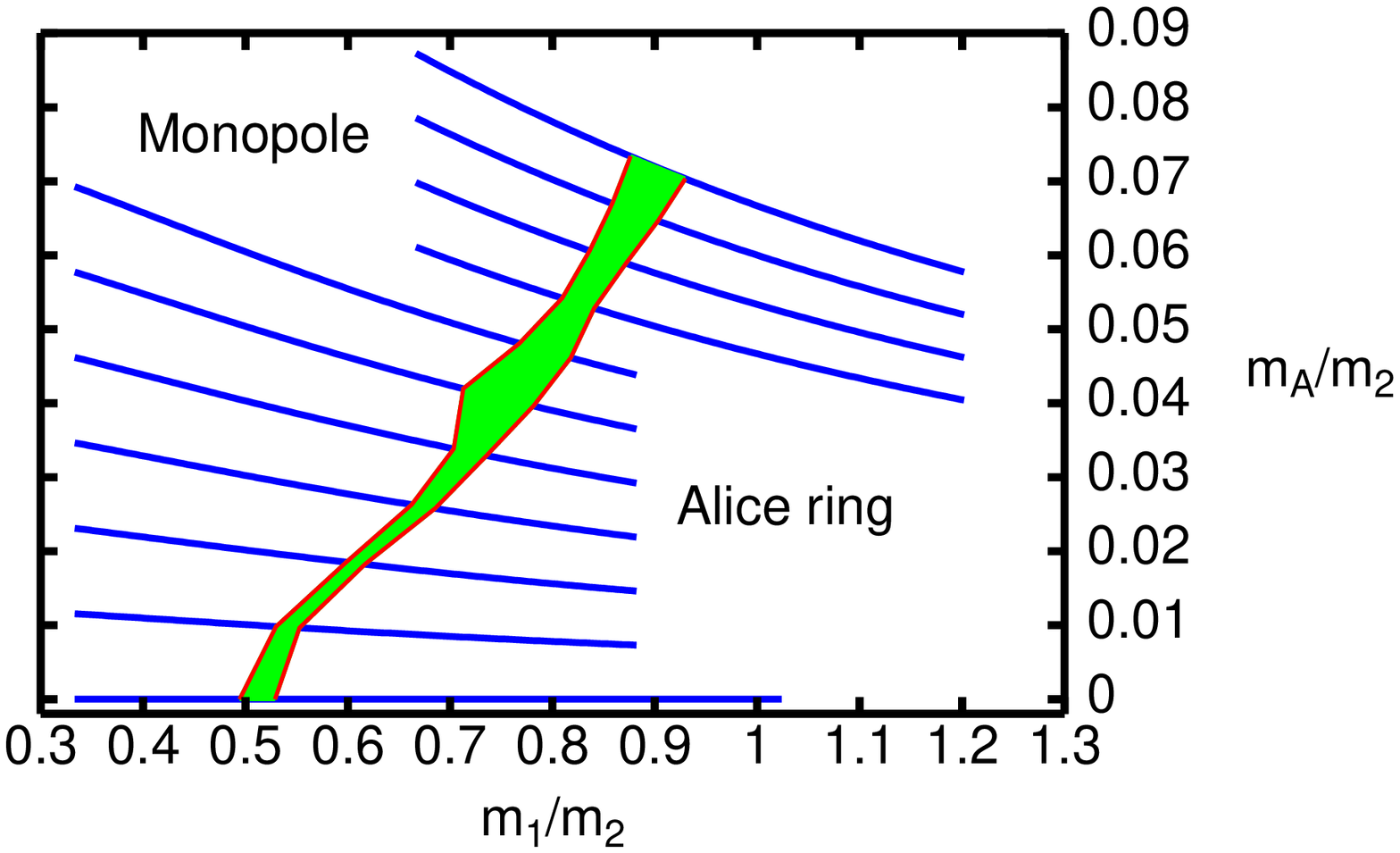,width=7.9cm,angle=270}}
\makebox[15.8cm][c]{\footnotesize{(c)}}
\caption[somethingelse]{\footnotesize These figures show the stability
results from the hysteresis type of experiments we did on the specific
lines in the parameter space of the model. The shaded regions give the
errors of the lines. Figure (a) shows the local instability line of
the alice ring configuration. Figure (b) shows the monopole
instability line and figure (c) shows the monopole meta-stability
line.}
\label{results}
\end{center}
\end{figure}

In figure \ref{results}(a-c) we give the results for the
meta-stability and instability lines. Figure \ref{results}(c) shows
the meta-stability line for the spherically symmetric magnetic
monopole solution. As we showed in section \ref{experiments} we can
use both lattice types to determine the monopole meta-stability line
as both lattice types move to this line from a different side as the
number of lattice points increase. We determined the monopole
meta-stability line up to the distance between the two monopole
meta-stability lines determined from the data of both lattice types at
the maximal number of lattice points we used. The plot shows the lines
on which we did the hysteresis type of experiments, the two monopole
meta-stability lines determined by the different lattice types and a
shaded region which is to represent the error in the position of the
monopole meta-stability line. Thus we did not extrapolate the results
from both types of lattices to an infinite number of lattice points we
just used the results from the lattices with the most lattice points
to corner the meta-stability line, see figure \ref{results}(c). This
line shows that the spherically symmetric magnetic monopole is not
always the lowest energy solution and cuts the parameter space into
two regions.

To determine the position of the instability line of the spherically
magnetic monopole solution, see figure \ref{results}(b), we just use
the results from the type (b) lattices as these show a clear point
where the monopole becomes unstable. This does mean we have to
extrapolate our results to a lattice with an infinite number of
points. For all lines in the parameter space we investigated we
observed that the change of the position, projected on the
$\frac{m_1}{m_2}$-axis, of this point from the 27$\times$27 to the
54$\times$54 lattice is about twice as big as the change in the
position from the 54$\times$54 to the 108$\times$108 lattice. We
estimate the position of the monopole instability point by
extrapolating this behavior to a lattice of an infinite number of
lattice points. This means that the estimated position of the
instability point lies at the same distance from the
108$\times$108-point as the 54$\times$54-point only on the other
side. The error we estimate as twice this distance. In figure
\ref{results}(b) we plotted these results. We plotted the lines on
which we did the hysteresis experiments. The shaded region is to
represent the error in the position of the monopole instability line
and we plotted the instability line obtained from the data of the
108$\times$108 lattices. The estimate of the instability line itself
is not plotted but is right in the middle of the shaded region.

In figure \ref{results}(a) we plotted the instability line of the
alice ring configuration. For most of the lines through the parameter
space we used the same technique as for the monopole meta-stability
line, for the rest we used the same principle as we used for the
monopole instability line.

Again we note that due to the fact that we used a variational approach
the monopole meta-stability and instability lines are upper bounds in
the sense that if less restrictions are forced upon the configurations
these lines can only move to the left, i.e., in favor of the cheshire
charged alice ring.

It is quite easy to understand why the alice ring is the lowest energy
configuration in the limit of large values of $\frac{m_1}{m_2}$ and
the monopole is the lowest energy solution in the limit of small
$\frac{m_1}{m_2}$. The two masses $m_1$ and $m_2$ correspond to energy
costs in deviations of the Higgs field from the vacuum
manifold. Deviations pure in the 'length' of the Higgs field
correspond to $m_1$. While deviations in the non-uniaxial direction
correspond to $m_2$. In the limit of $\frac{m_1}{m_2}\to
0$\footnote{Note that this limit can in fact not be taken as the
minimum value of $\frac{m_1}{m_2}$ at which the broken vacuum is still
the true vacuum is equal to $\frac{1}{3}$. For smaller values of
$\frac{m_1}{m_2}$ the unbroken vacuum is the true vacuum. We come back
to this point in the conclusions and outlook section.} the
non-uniaxial deviations are suppressed and the spherically symmetric
(uniaxial) magnetic monopole is the lowest energy solution. In the
limit of $\frac{m_1}{m_2}\to\infty$ one would expect an 'escape' in
the non-uniaxial direction. This signals the meta-stability of the
monopole and implies that the cheshire charged alice ring is the
lowest energy solution. In this case the length of the Higgs field
never becomes zero as can bee seen in figure \ref{defects}(b), i.e.,
the quantity $\left(1-\frac{Tr\Phi^2}{6f^2}\right)$ never becomes
equal to one.

We pointed out that one of the main factors determining the monopole
core meta-stability is the mass ratio $\frac{m_1}{m_2}$ of the charged
and neutral Higgs particles. As the mass of the charged excitation
becomes much smaller than the mass of the neutral excitation the
't~Hooft Polyakov magnetic monopole is expected to become
meta-stable. Clearly this argument holds generically and one expects a
core meta-/instability to be a general feature of the 't~Hooft
Polyakov magnetic monopole in models with charged Higgs
excitations. The nice thing of alice type models is that they
naturally suggest an alternative configuration to the 't Hooft
Polyakov type monopoles: the magnetically cheshire charged alice loop.

\section{Conclusions and outlook}
In this paper we investigated the core structure of the unit charge
magnetic monopole and discussed the numerical methods we employed in
some detail. We already showed in a previous letter \cite{jelper2}
that the core structure of the magnetic monopole is not necessarily
spherically symmetric. The model has three mass scales, two of them
refer to the Higgs field, one to its length and the other to deviation
from the uniaxial direction. The third mass scale is set by the mass
of the broken gauge fields. The topologically non-trivial boundary
conditions can be met by an ``escape'' in a non-uniaxial
direction. This possibility allows for the length of the Higgs field
to stay finite in the core and not go to zero as would be necessary
otherwise. As the ratio of the masses, $\frac{m_1}{m_2}$, increases it
becomes harder energetically to decrease the length of the Higgs field
and one would expect an escape in the non-uniaxial direction.

At the end of section \ref{someresults} we argued that a core
instability of the `t~Hooft Polyakov magnetic monopole is a general
feature of models with charged excitations of the Higgs field. The
instability occurs in the region of the parameter space where the
charged excitations are much lighter than the neutral excitations. In
alice electrodynamics there is also an other, somewhat independent
motivation to question the core stability of the spherically symmetric
magnetic monopole, provided by the possibility of an alice ring
which can carry a magnetic cheshire charge.

Within our ansatz we determined the meta- and instability regions of
the spherically symmetric magnetic monopole. We also found that, as
expected, the competing configuration is the magnetically cheshire
charged alice ring. We did also stumble upon the somewhat unwanted
splitcore configurations but fortunately they never became the lowest
energy solutions. As we used a variational approach we cannot claim
that the alice ring configurations we found are exact solutions to the
equations of motion. However they do have every feature which one
would expect from an exact alice ring solution. Also, because we used
a variational approach the regions of meta-stability and
global-stability of the spherically magnetic monopole are with respect
to energetic upper bounds so that with respect to the exact solutions,
these regions can only become smaller.

As a final comment we want to come back to the fact that the minimum
value of $\frac{m_1}{m_2}=\frac{1}{3}$ for which the broken vacuum is
the true vacuum. In section \ref{someresults} we gave a simple
explanation of why the spherically magnetic monopole becomes globally
unstable in the limit of large values of $\frac{m_1}{m_2}$. In the
opposite limit one would expect the monopole to be the spherically
magnetic monopole to be the global stable solution. However as the
minimum value of $\frac{m_1}{m_2}=\frac{1}{3}$ there is no guarantee
that this ever happens. It could just be that in this or similar
models the `t~Hooft Polyakov type magnetic monopole is never globally
stable.

Acknowledgment: This work was partially supported by the ESF COSLAB
program.


\end{document}